\documentclass[8.5pt,twoside,twocolumn,final]{article}
\usepackage[super,sort&compress,comma]{natbib} 
\usepackage{mhchem}
\usepackage{times,mathptmx}
\usepackage{sectsty}
\usepackage{balance} 

\usepackage{graphicx} 
\usepackage{lastpage}
\usepackage[format=plain,justification=raggedright,singlelinecheck=false,font=small,labelfont=bf,labelsep=space]{caption} 
\usepackage{fancyhdr}
\usepackage[utf8]{inputenc}
\usepackage{color}
\usepackage{xspace}
\usepackage{upgreek}
\usepackage{subfig}
\usepackage{url}

\newcommand{\micron}{\ensuremath{\upmu \text{m}}}
\newcommand{\Dcrit}{\ensuremath{D_\mathrm{cr}}\xspace}
\newcommand{\mods}{\kappa_\mathrm{s}}
\newcommand{\modal}{\kappa_\alpha}
\newcommand{\modb}{\kappa_\mathrm{b}}
\newcommand{\Canum}{\ensuremath{Ca}\xspace}
\newcommand{\Renum}{\ensuremath{Re}\xspace}
\newcommand{\dlat}{\ensuremath{D_\perp}\xspace}
\newcommand{\ie}{\emph{i.e.}\xspace}
\newcommand{\eg}{\emph{e.g.}\xspace}
\newcommand{\via}{\emph{via}\xspace}

\begin{document}

\thispagestyle{plain}
\fancypagestyle{plain}{
\fancyhead[L]{\includegraphics[height=8pt]{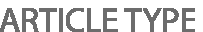}}
\fancyhead[C]{\hspace{-1cm}\includegraphics[height=20pt]{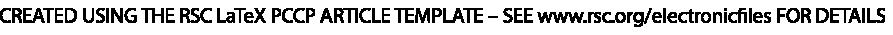}}
\fancyhead[R]{\includegraphics[height=10pt]{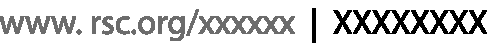}\vspace{-0.2cm}}
\renewcommand{\headrulewidth}{1pt}}
\renewcommand{\thefootnote}{\fnsymbol{footnote}}
\renewcommand\footnoterule{\vspace*{1pt}%
\hrule width 3.4in height 0.4pt \vspace*{5pt}} 
\setcounter{secnumdepth}{5}

\makeatletter 
\def\subsubsection{\@startsection{subsubsection}{3}{10pt}{-1.25ex plus -1ex minus -.1ex}{0ex plus 0ex}{\normalsize\bf}} 
\def\paragraph{\@startsection{paragraph}{4}{10pt}{-1.25ex plus -1ex minus -.1ex}{0ex plus 0ex}{\normalsize\textit}} 
\renewcommand\@biblabel[1]{#1}            
\renewcommand\@makefntext[1]%
{\noindent\makebox[0pt][r]{\@thefnmark\,}#1}
\makeatother 
\renewcommand{\figurename}{\small{Fig.}~}
\sectionfont{\large}
\subsectionfont{\normalsize} 

\fancyfoot{}
\fancyfoot[LO,RE]{\vspace{-7pt}\includegraphics[height=9pt]{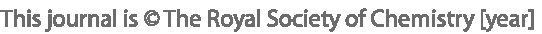}}
\fancyfoot[CO]{\vspace{-7.2pt}\hspace{12.2cm}\includegraphics{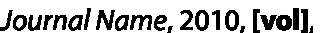}}
\fancyfoot[CE]{\vspace{-7.5pt}\hspace{-13.5cm}\includegraphics{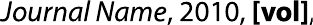}}
\fancyfoot[RO]{\footnotesize{\sffamily{1--\pageref{LastPage} ~\textbar  \hspace{2pt}\thepage}}}
\fancyfoot[LE]{\footnotesize{\sffamily{\thepage~\textbar\hspace{3.45cm} 1--\pageref{LastPage}}}}
\fancyhead{}
\renewcommand{\headrulewidth}{1pt} 
\renewcommand{\footrulewidth}{1pt}
\setlength{\arrayrulewidth}{1pt}
\setlength{\columnsep}{6.5mm}
\setlength\bibsep{1pt}

\twocolumn[
  \begin{@twocolumnfalse}
\noindent\LARGE{\textbf{Deformability-based red blood cell separation in deterministic lateral displacement devices --- a simulation study}}
\vspace{0.6cm}

\noindent\large{\textbf{Timm Krueger,$^{\ast}$\textit{$^{a,b}$} David Holmes,\textit{$^{c,d}$} and
Peter V. Coveney\textit{$^{b}$}}}\vspace{0.5cm}

\noindent\textit{\small{\textbf{Received Xth XXXXXXXXXX 20XX, Accepted Xth XXXXXXXXX 20XX\newline
First published on the web Xth XXXXXXXXXX 200X}}}

\noindent \textbf{\small{DOI: 10.1039/b000000x}}
\vspace{0.6cm}

\noindent \normalsize{%
We show, \via three-dimensional immersed-boundary-finite-element-lattice-Boltzmann simulations, that deformability-based red blood cell (RBC) separation in deterministic lateral displacement (DLD) devices is possible.
This is due to the deformability-dependent lateral extension of RBCs and enables us to predict \emph{a priori} which RBCs will be displaced in a given DLD geometry.
Several diseases affect the deformability of human cells.
Malaria-infected RBCs or sickle cells, for example, tend to become stiffer than their healthy counterparts.
It is therefore desirable to design microfluidic devices which can detect those diseases based on the cells' deformability fingerprint, rather than preparing samples using expensive and time-consuming biochemical preparation steps.
Our findings should be helpful in the development of new methods for sorting cells and particles by deformability.
}
\vspace{0.5cm}
 \end{@twocolumnfalse}
  ]

\footnotetext{\textit{$^{a}$~Institute for Materials and Processes, School of Engineering, University of Edinburgh, The King's Buildings, Edinburgh EH9 3JL, Scotland, United Kingdom; Tel: +44 131 650 5679; E-mail: timm.krueger@ed.ac.uk}}
\footnotetext{\textit{$^{b}$~Centre for Computational Science, University College London, 20 Gordon Street, London WC1H 0AJ, United Kingdom.}}
\footnotetext{\textit{$^{c}$~Sphere Fluidics Ltd., The Jonas Webb Building, Babraham Research Campus, Babraham, Cambridge CB22 3AT, United Kingdom.}}
\footnotetext{\textit{$^{d}$~London Centre for Nanotechnology, University College London, 17-19 Gordon Street, London WC1H 0AH, United Kingdom.}}

\section{Introduction}

Particle separation is important for a wide range of clinical and analytical processes such as the analysis of unknown suspensions of biological particles.
Blood, for example, is a complex mixture, comprised of cells with widely differing function, size and mechanical properties; these include platelets, erythrocytes (red blood cells, RBCs) and leukocytes (\eg neutrophils, lymphocytes, monocytes, basophils, eosinophils).
Depending on physiological state or condition, all these cells can undergo distinct morphological alterations.
These changes in a cell's biophysical or biomechanical properties can arise as a result of a vast array of biological, chemical and physical stimuli.
Changes in cell deformability are known to influence many disease pathologies.

Metastatic cancer cells have been shown to have a ``softer'' phenotype to that of their healthy counterparts.~\cite{guck_optical_2005}
Diseases such as sickle cell anemia, and malaria,~\cite{suresh_connections_2005} as well as hereditary blood disorders exhibit erythrocyte ``stiffening'' with onset and progression of the pathological state.
In malaria (\eg \emph{Plasmodium falciparum} infection), membrane stiffness of the parasitised erythrocytes has been shown to increase by almost two orders of magnitude as the intracellular parasite matures.~\cite{quinn_combined_2011}

Conventional cell separation strategies typically rely on intrinsic properties of the cells (\eg density) or external labels to distinguish between cell types.
For example, fluorescence-activated cell sorting (FACS) and magnetic-activated cell sorting (MACS) use fluorescent or magnetic conjugated antibodies in order to label the cells of interests; and either trigger an active sorting process or sort passively \via interaction of the label with an applied force field.
Centrifugation allows separation of blood into its cellular components based on the differential densities of the different cell types.
These methods have been refined over the years and allow for very high resolution sorting.

Microfluidic separation methods\cite{pamme_continuous_2007, gossett_label-free_2010, lenshof_continuous_2010, karimi_hydrodynamic_2013} allow for novel sorting modalities and can take advantage of smaller scales of microfluidic devices (\ie dimensions similar to that of the cells).
As such, a number of label-free microfluidic platforms have been proposed that use non-traditional biomarkers; these include cell size, shape, electrical properties (\eg polarisability), density, deformability and hydrodynamic properties.

\begin{figure}
 \centering
 \includegraphics[width=0.48\linewidth]{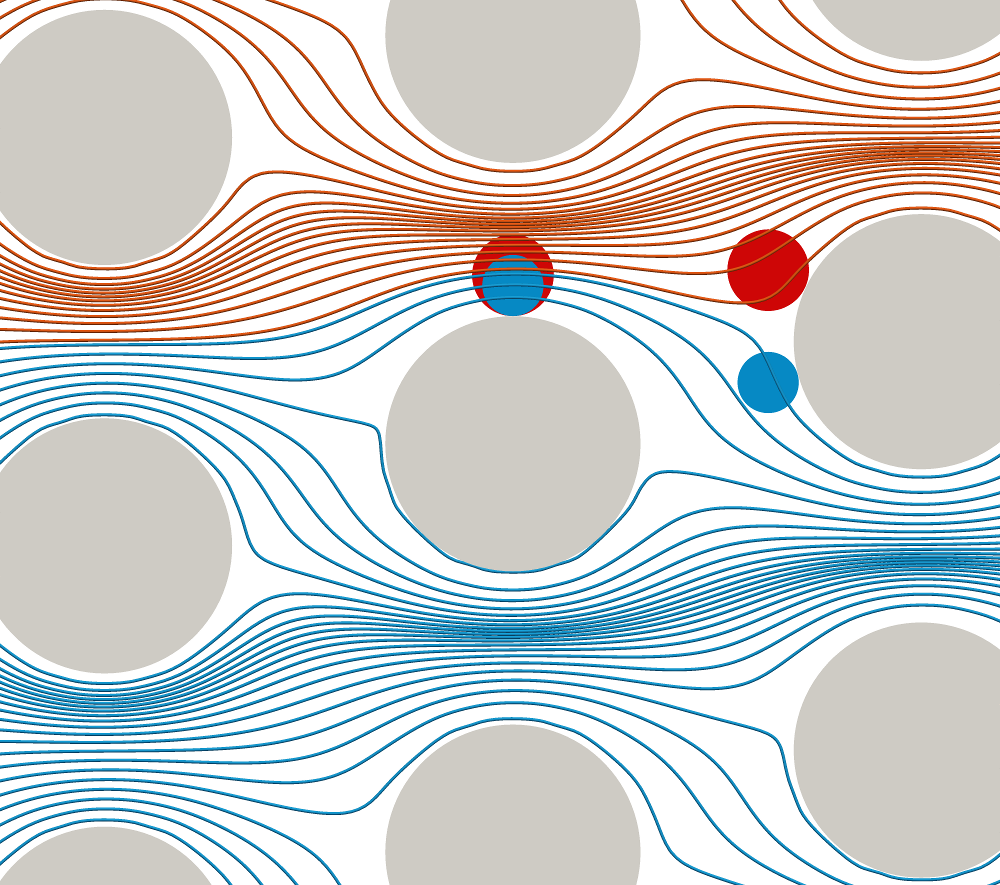}
 \caption{\label{fig:geometry1} (Colour online) Geometry of our DLD setup. An array of micron-sized pillars is set at an angle to the direction of fluid flow (from left to right) through a microfluidic channel. As described previously,~\cite{inglis_critical_2006} the total fluid flux (shown as streamlines) through the pillar gap can be divided into a number of flow streams, each carrying equal fluid flux. Depending on their size, particles flowing in the device follow different paths. Particles (blue) which are smaller than a critical diameter \Dcrit can pass the downstream obstacle on the right, while larger particles (red) are forced to follow another set of streamlines and pass the downstream obstacle on the left. The larger particles are successively displaced while the smaller ones essentially move horizontally on zigzag trajectories.}
\end{figure}

Only ten years ago, \citet{huang_continuous_2004} proposed a simple and effective microfluidic design to separate particles by size: deterministic lateral displacement (DLD).
These devices have been used for label-free particle sorting, \ie separation based on intrinsic properties rather than (bio)chemical pretreatment or external fields.
DLD devices consist of arrays of pillars positioned within a flow channel and are capable of high-resolution continuous sorting of cells and other microscopic particles.
The underlying idea is the steric interaction between the flowing particles and those pillars (Fig.~\ref{fig:geometry1}).
Depending on their size, some particles collide with the obstacles, which leads to irreversible trajectories and cross-streamline motion, even in Stokes flow.
Objects smaller than a critical size \Dcrit move in the direction of flow (\ie along the length of the channel), while objects larger than the critical size move in a direction defined by the pillar arrangement (\ie they are laterally displaced).
Different particle species can therefore be collected at different outlets of the device.
As the mechanism behind DLD is based on deterministic (non-diffusive) separation, it can be scaled up to higher flow rates.~\cite{davis_deterministic_2006}

Several groups investigated how to improve the performance of DLD devices.
Clogging is always a problem for most mechanical filtration methods because the typical particle size is usually of the order of the gaps between obstacles.
DLD devices typically work well with a gap sizes three times that of the largest particles, thus reducing issues related to clogging.
\Citet{loutherback_deterministic_2009} proposed a combination of triangular posts and an oscillating flow and later~\cite{loutherback_improved_2010} showed that equilateral triangular pillars with sharp vertices (rather than polygons with more vertices, rounded triangluar or even circular pillars) improve the performance due to their enhanced ability to separate particles of a defined size using devices with larger gap sizes.

High throughput is generally desired as it permit the analysis of more fluid volume in less time.
This can be achieved by parallelisation or increasing the flow rate.
\Citet{loutherback_deterministic_2012} and later \citet{liu_rapid_2013} performed high-throughput experiments showing that cancer cell separation from diluted whole blood is possible at flow rates of up to several $\mathrm{ml} / \mathrm{min}$.
However, increasing the flow rate leads to larger viscous stresses which in turn affect the behaviour of biological objects which are often compliant.
While DLD devices function in a straightforward manner for rigid spherical particles, deformability and non-sphericity of such objects can have a large influence on the particle trajectories within these devices.

\citet{davis_deterministic_2006} were the first to use DLD devices to separate blood components by size.
The authors briefly addressed the issue of possible particle deformation due to local stresses but did not investigate this further.
\Citet{holm_separation_2011} used DLD to separate RBCs from \emph{Trypanosoma cyclops} parasites.
\Citet{inglis_scaling_2011} proposed parallelised DLD devices to enrich leukocytes in undiluted whole blood, but did not consider blood cell deformation.
\Citet{al-fandi_new_2011} emphasised that deformable and non-spherical particles are usually more difficult to separate in DLD devices since their trajectories are irratic due to flow-induced deformation and tumbling.
In order to minimise these undesired effects, the authors proposed novel pillar shapes such as diamond or airfoil.

Although \citet{quek_separation_2011} performed 2D simulations of initially circular deformable particles and \citet{beech_sorting_2012} proposed particle deformability as an additional marker for separation, the authors did not investigate the role of particle deformability in DLD devices systematically.
\Citet{inglis_determining_2008} mentioned the potential impact of leukocyte deformability on their apparent size in DLD devices.

Rather than avoiding or ignoring RBC deformability, we aim to understand the effect of deformability on the RBC behaviour in DLD devices.
To this end we have developed (section~\ref{sec:method}) and benchmarked (section~\ref{sec:simulations}) a model for deformable RBCs in DLD devices.
In this study, we have restricted ourselves to one free DLD geometry parameter, the row shift $d$ (Fig.~\ref{fig:geometry2}).
The RBC deformability is described in terms of the capillary number \Canum which is the ratio of viscous, deforming stresses to the intrinsic, restoring stress of an RBC.
Our analysis of the deformation characteristics of a single RBC during its passage (section~\ref{sec:results}) shows that deformability-based RBC separation in a DLD device is possible.
We present a \Canum-$d$ diagram (``phase space'') for the particle trajectories and rationalise the separation characteristics in terms of the apparent lateral RBC diameter.
We further suggest a way to predict the observed trajectory type (displaced or undisplaced) based only on properties of the RBC and the known critical separation diameter of the device.

Our work should contribute to the development of future DLD devices for deformability-based particle separation.
This could be applied, for example, to the mechanical detection of malaria-infected RBCs.

\section{Numerical method and geometry}
\label{sec:method}

In the following we briefly present the employed numerical model (section \ref{sec:model}) and define the chosen flow geometry and parameters (section \ref{sec:geometry}).

\subsection{Numerical model}
\label{sec:model}

\begin{figure}
 \centering
 \includegraphics[width=0.48\linewidth]{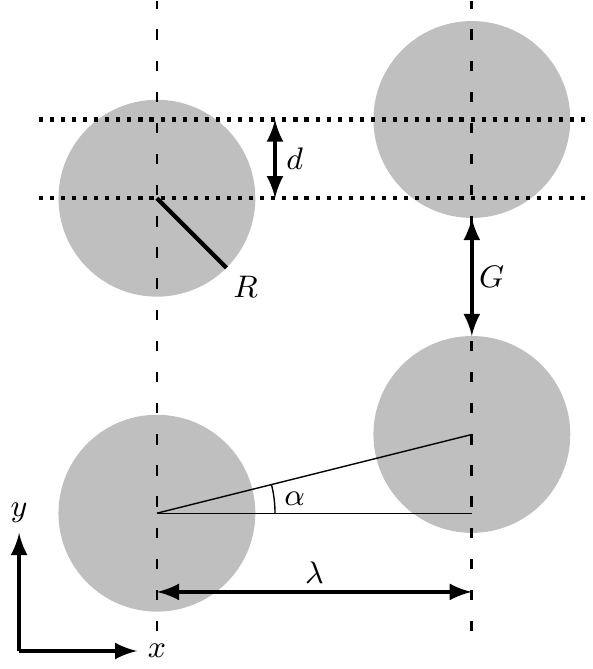} \\
 \caption{\label{fig:geometry2} Our DLD geometry. The geometry is defined by the pillar radius $R$, the centre-to-centre distance $\lambda$ and the row shift $d$. The gap size $G$ and shift angle $\alpha$ are then uniquely determined. The critical diameter \Dcrit is a function of these geometrical parameters.}
\end{figure}

We employ a combination of lattice-Boltzmann method (LBM)~\cite{succi_lattice_2001, aidun_lattice-boltzmann_2010} for the fluid phase, finite-element method (FEM)~\cite{charrier_free_1989, shrivastava_large_1993} for the membrane dynamics and immersed-boundary method (IBM)~\cite{peskin_immersed_2002} for the bidirectional fluid-membrane coupling as previously described and benchmarked.~\cite{kruger_efficient_2011, kruger_crossover_2013, gross_rheology_2014}
The flow is driven by a pressure gradient along the $x$-axis, mimicked by a constant force density.
The RBC membranes are characterised by three elastic moduli: shear elasticity $\mods$, local area dilation resistance $\modal$ and bending modulus $\modb$.
Additionally, we have implemented a viscosity contrast between the RBC interior and exterior regions.~\cite{frijters_parallelised_2014} The interior and exterior viscosities are $\eta = \frac{5}{6}$ and $\frac{1}{6}$ in lattice units, respectively.
The numerical fluid density is unity in the entire domain.

We use the bounce-back boundary condition~\cite{ladd_numerical_1994} to describe the confining walls and the obstacles in the simulated DLD geometry.
The confining walls are located at $z = \pm H / 2$ where $H$ is the height of the DLD device.
A short-range repulsion is applied between the RBC and the obstacle surfaces to avoid overlap.
RBC surface nodes which come closer than one lattice constant to the obstacle are repelled, which leads to a thin lubrication layer.

In order to reduce the simulation domain to a single obstacle unit cell, we use shifted periodicity conditions in the flow direction.
Fluid and particles leaving the computational domain along the flow axis ($x$-axis) re-enter the domain from the other side but are displaced by the row shift $d$ (Fig.~\ref{fig:geometry2}) along the $y$-direction.
We only allow integer shifts in lattice units to avoid additional interpolations.
The flow is periodic along the $y$-axis.
Fig.~\ref{fig:geometry2} therefore shows four identical replications of the same unit cell, each containing only one obstacle.

The present model is athermal with no fluctuations or intrinsic cell diffusion.
The Péclet number is, therefore, infinite and the flow is fully deterministic.
In realistic DLD devices the Péclet number is always finite and thermal diffusion may reduce the separation efficiency of very small particles.~\cite{davis_deterministic_2006}
RBCs, however, are sufficiently large that thermal diffusion is negligible in most applications.

\subsection{Geometry and parameters}
\label{sec:geometry}

The chosen geometry is shown in Fig.~\ref{fig:geometry2}.
The DLD device is characterised by pillars with a circular cross-section (radius $R$) and a centre-to-centre distance $\lambda$ (both along the $x$- and the $y$-axis).
This results in a gap size $G = \lambda - 2 R$.
Neighbouring pillar rows are shifted by a displacement (row shift) $d$, which in turn defines the dimensionless displacement parameter $\epsilon = d / \lambda$ and the shift angle $\alpha$ \via $\tan \alpha = \epsilon$.

In our simulations we have fixed $R = 10\, \micron$, $\lambda = 32\, \micron$ and therefore $G = 12\, \micron$.
Moreover, the device depth along the $z$-direction is $H = 4.8\, \micron$.
This way, RBCs are forced to move parallel to the confining bottom and top walls.
The row shift $d$ is the only free geometrical parameter in the present study.
Thus, for rigid spheres, the critical diameter \Dcrit is a function of $d$ only.~\cite{holm_separation_2011}

\begin{figure}
 \centering
 \includegraphics[width=0.8\linewidth]{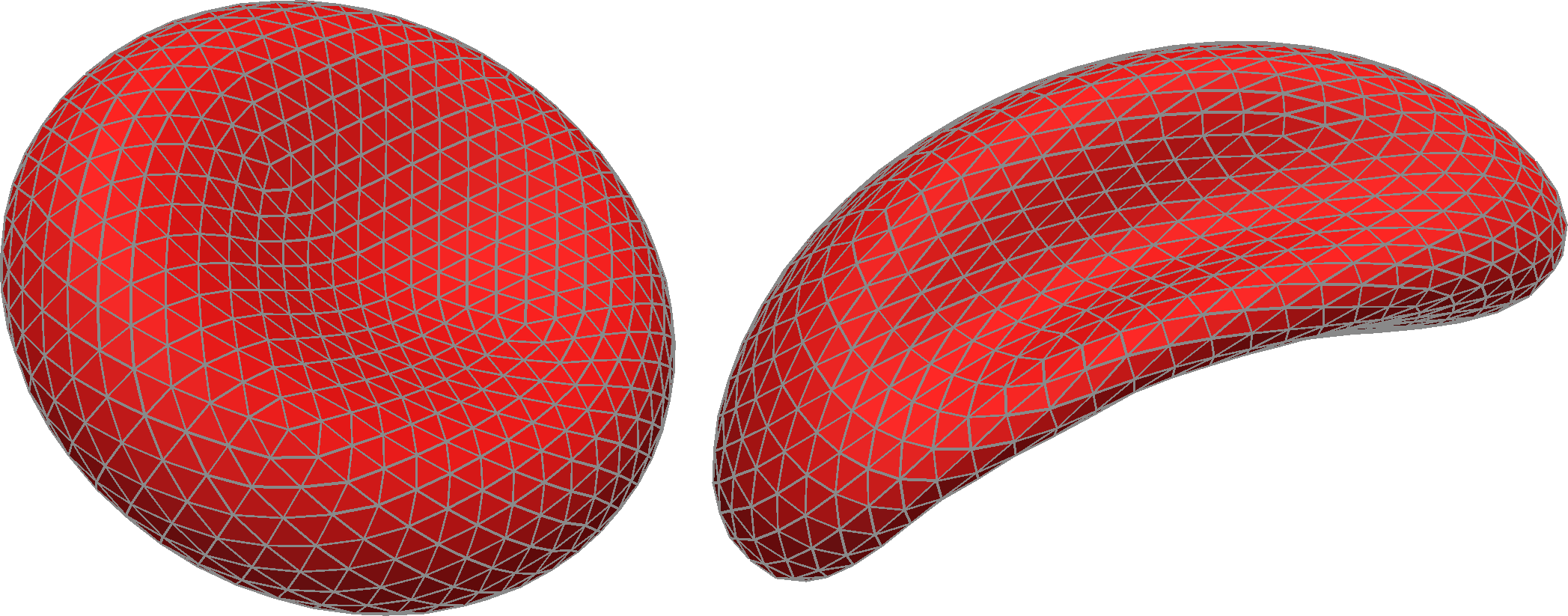} \\
 \caption{\label{fig:rbc_geometry} (Colour online) Mesh of an undeformed (left) and deformed (right) RBC. The cell surface is approximated by $2\,000$ flat triangular elements.}
\end{figure}

The shape of the modelled RBCs corresponds to the parametrisation first reported by \citet{evans_improved_1972}.
All RBCs in our simulations have the same in-plane radius, $r = 3.9\, \micron$, and we use $2\,000$ triangular elements to describe the RBC surface as shown in Fig.~\ref{fig:rbc_geometry}.
The RBC is assumed to be stress-free in its equilibrium state.

It is convenient to introduce a dimensionless parameter denoting the RBC deformability.
Since cell deformation in Stokes flow is caused by viscous stresses, the capillary number, which is one ratio of viscous stress to characteristic membrane stress, is a suitable parameter.
The latter is defined by $\mods / r$ where $\mods$ is the in-plane shear elasticity.
The typical viscous stress is given by the applied pressure gradient $p'$ and the gap length scale $\ell := \sqrt{G H}$ as geometric average of gap width and height.
We therefore define the capillary number as
\begin{equation}
 \Canum = \frac{p' \ell r}{\mods}.
\end{equation}

Inertial effects in DLD devices are usually negligible and the Reynolds number
\begin{equation}
 \Renum = \frac{\bar u \ell}{\nu},
\end{equation}
defined \via the average flow velocity $\bar u$ and the kinematic viscosity $\nu$, is of the order of $10^{-2}$ (except for some high-throughput experiments with \Renum up to 40).~\cite{loutherback_deterministic_2012}
We will therefore not consider \Renum as an additional free parameter; rather we will demonstrate in section \ref{sec:simulations_stokes} that the chosen simulation parameters correspond to Stokes flow.

\section{Simulations}
\label{sec:simulations}

In this section we describe the simulation setup and the results of the benchmark tests. We have used $\Delta x = 0.4\, \micron$ as the lattice constant for all simulations in this work. In particular, the size of a DLD array unit cell is $80 \Delta x \times 80 \Delta x \times 12 \Delta x$, and the RBC diameter is $2 r = 19.5\, \Delta x$.

\subsection{Stokes flow assumption and flow resistance}
\label{sec:simulations_stokes}

We have run a series of simulations without particles to validate the Stokes flow assumption which predicts the following observations: (i) The average flow velocity $\bar u$ should be strictly proportional to the applied pressure gradient $p'$. (ii) The shape of the streamlines should be the same when the geometry is mirrored: $\mathbf{x} \to - \mathbf{x}$.

\begin{figure}
 \centering
 \includegraphics[width=\linewidth]{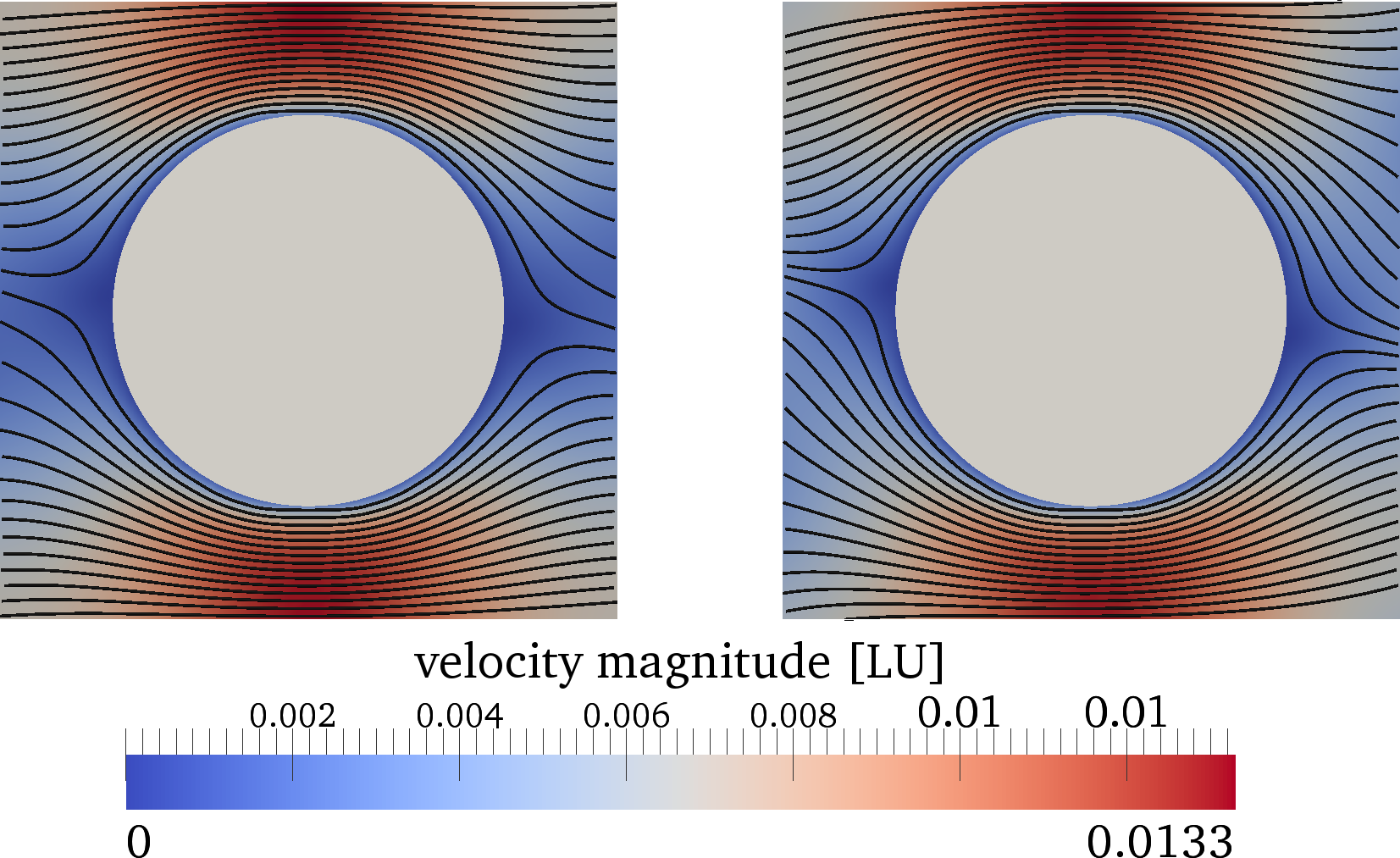} \\
 \caption{\label{fig:streamlines} (Colour online) Streamlines and colour-coded velocity magnitude (in lattice units) half-way between bottom and top walls in one unit cell of the pillar array for $\epsilon = 5/80$ (left) and $\epsilon = 20/80$ (right) in the absence of particles at $\Renum = 0.76$. The streamlines are virtually identical when the flow direction is reversed.}
\end{figure}

The tested parameters are $\epsilon = 5/80$, $13/80$ and $20/80$ (corresponding to $d = 2.0$, $5.2$ and $8.0\, \micron$), each for $\Renum = 0.076$ and $0.76$. This is achieved by setting the numerical viscosity to $\nu = \tfrac{1}{6}$ and the force density to $9.36 \times 10^{-6}$ and $9.36 \times 10^{-5}$, respectively (all quantities in lattice units). For all investigated values of $\epsilon$ we have found that the ratio $\bar u / p'$ increases by less than $3 \times 10^{-5}$ upon a tenfold increase of \Renum. This indicates that, in the current parameter range, the Reynolds number is irrelevant. This interpretation is also supported by the shape of the streamlines (\cf Fig.~\ref{fig:streamlines}) which are virtually identical under inversion of the coordinate system. We therefore choose $\Renum = 0.76$ for all subsequent simulations, which allows us to achieve shorter simulation times since the time step obeys $\Delta t \propto \Renum$ and the total simulation runtime is thus proportional to $\Renum^{-1}$.

Furthermore we define the flow resistance $\varrho$ \via $\bar u = p' / \varrho$. It denotes the required pressure gradient to achieve a desired average flow velocity. From the simulations we find that $\varrho = 8.8 \times 10^8\, \tfrac{\mathrm{Pa}}{\mathrm{m}} / \tfrac{\mathrm{m}}{\mathrm{s}} = 8.8\, \tfrac{\mathrm{bar}}{\mathrm{m}} / \tfrac{\mathrm{mm}}{\mathrm{s}}$. This value is the same for all tested displacements $\epsilon$ up to $1\%$. For a typical device length of about $30\, \mathrm{mm}$ and an average flow velocity of $1\, \mathrm{mm}/\mathrm{s}$, one would therefore require a pressure drop of about $0.27\, \mathrm{bar}$ along the pillar array. Note that this result is only valid for the chosen geometry, in particular for a device depth of $H = 4.8\, \micron$. Larger depths are usually desired to increase the throughput, but this would give the RBCs additional rotational degrees of freedom as they would not longer be in forced alignment with the bottom and top walls.~\cite{beech_sorting_2012}

\subsection{Critical diameters}

We have analysed the flow streamlines in our simulations without particles and obtained the streamline separation distance $s$ in a plane parallel to and half-way between the walls.
$s$ is defined as the shortest distance between a pillar and the streamline which ends in a stagnation point in the next downstream pillar row.
Since the streamline density is highest in the region between two laterally neighbouring pillars (Fig.~\ref{fig:streamlines}), the point of closest proximity is located somewhere on the line connecting those two pillars (vertical dashed lines in Fig.~\ref{fig:geometry2}).
We then assume that the critical diameter \Dcrit of a rigid sphere is twice this separation distance $s$: $\Dcrit = 2 s$.

The results obtained for $\Dcrit(\epsilon)$ as listed in Table~\ref{tab:streamlines} can be accurately captured by the simple scaling law
\begin{equation}
 \label{eq:separation_law}
 \Dcrit(\epsilon) = 2 s(\epsilon) = 19.2\, \micron \times \epsilon^{0.76}
\end{equation}
in the range $\epsilon \in [5/80, 22/80]$.
Note that the prefactor and exponent will generally depend on the details of the system, \eg the pillar shape and system depth.

\begin{table}
 \small
 \caption{\label{tab:streamlines} Streamline separation distances $s$ as function of row shift $d$ as predicted by simulations without particles. These data can be approximated by Eq.~\eqref{eq:separation_law}.}
 \begin{tabular*}{0.5\textwidth}{@{\extracolsep{\fill}}lll}
  \hline
  row shift & displacement parameter & separation distance \\
  $d\, [\micron]$ & $\epsilon$ & $s\, [\micron]$ \\
  \hline
  2.0 & $5/80$ & 1.2 \\
  3.2 & $8/80$ & 1.7 \\
  4.4 & $11/80$ & 2.1 \\
  5.2 & $13/80$ & 2.4 \\
  6.0 & $15/80$ & 2.7 \\
  6.8 & $17/80$ & 3.0 \\
  8.0 & $20/80$ & 3.4 \\
  8.8 & $22/80$ & 3.6 \\
  \hline
 \end{tabular*}
\end{table}

\subsection{RBC mechanics benchmark}

\begin{figure}
 \centering
 \subfloat[relaxed and stretched RBC ($F = 42\, \mathrm{pN}$)]{\includegraphics[width=0.475\linewidth]{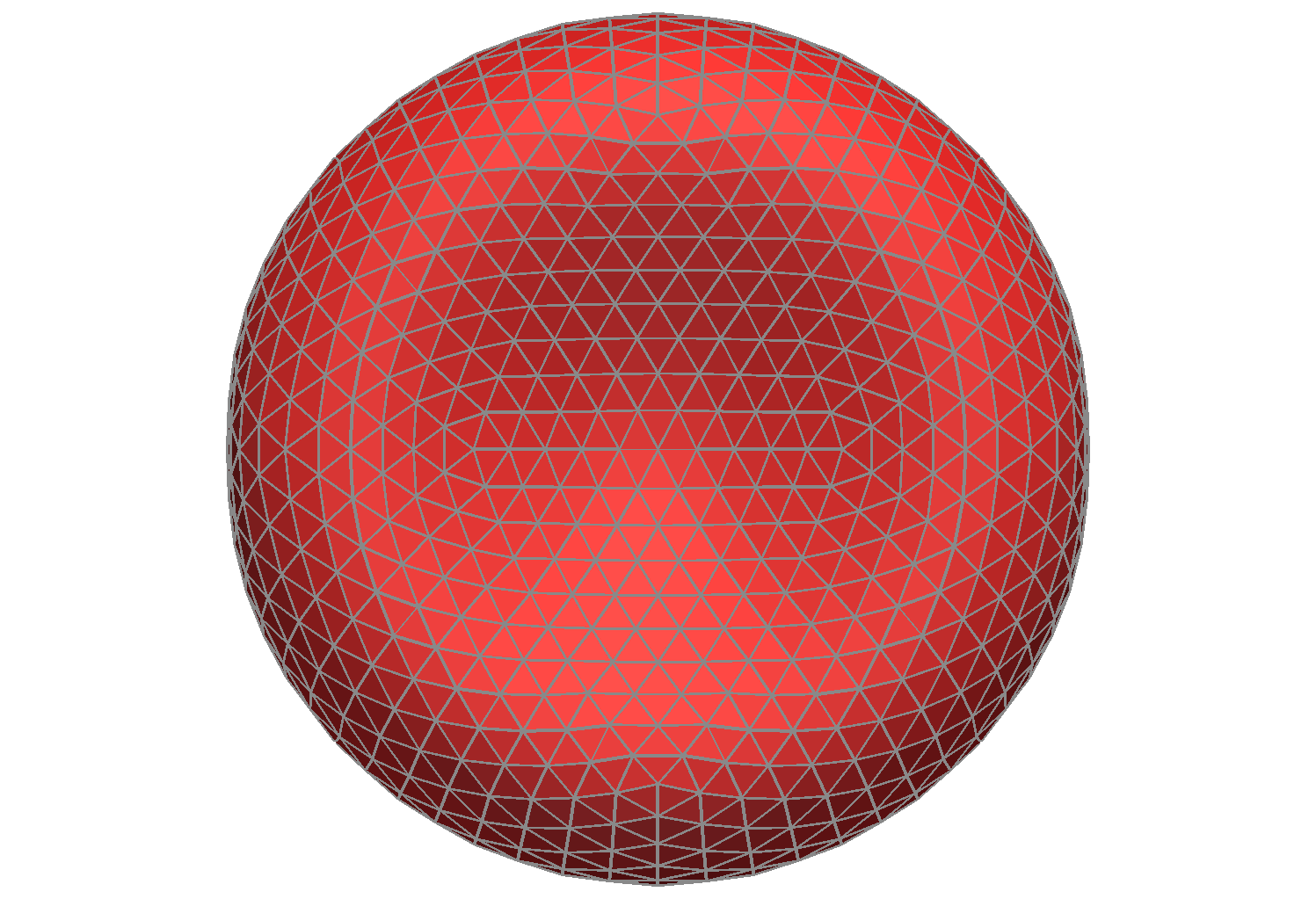} \hfill
 \includegraphics[width=0.475\linewidth]{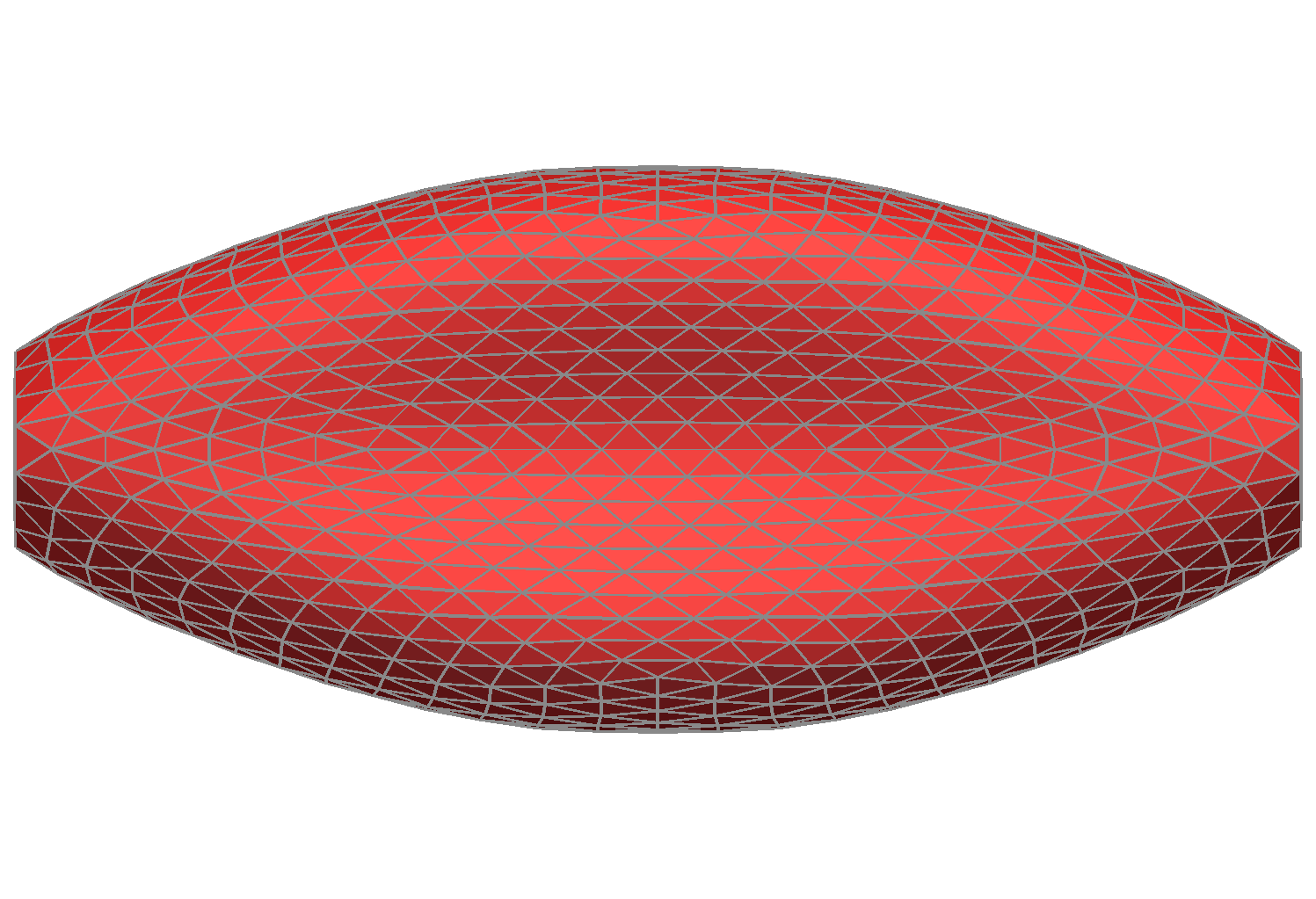}} \\
 \subfloat[\label{fig:tweezer_data} simulated (sim) and experimental (exp) stretching diameters]{\includegraphics[width=\linewidth]{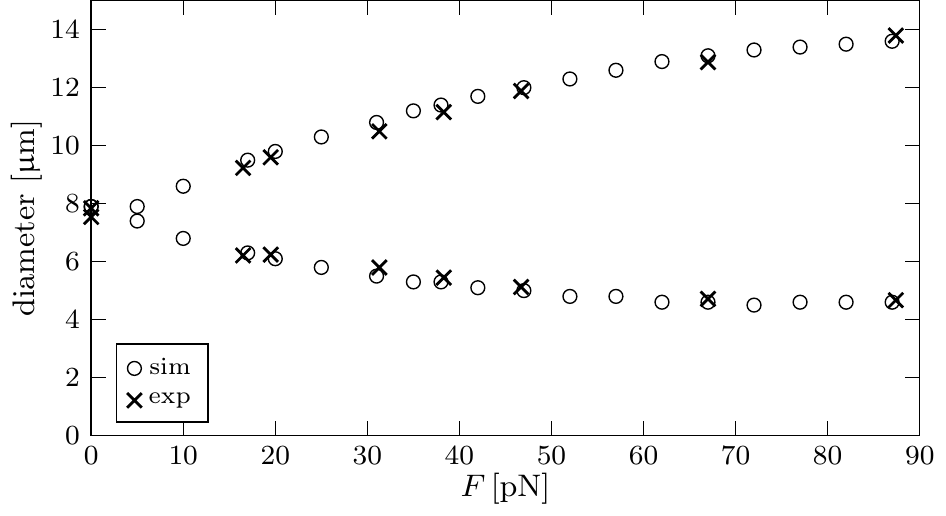}} \\
 \caption{\label{fig:tweezer} (Colour online) An RBC is stretched by applying a force $F$ to opposite ends. The radius of the force contact area, which can be clearly seen in (a), is $1\, \micron$. The stationary longitudinal and transversal diameters are shown in (b) as a function of the applied force $F$. Experimental data for healthy RBCs have been taken from Fig.~7 in \citet{suresh_connections_2005}}
\end{figure}

In order to check the validity of the RBC model we have simulated the stretching of an RBC in an optical tweezer.~\cite{suresh_connections_2005}
Forces of equal magnitude and opposite direction are applied to the ends of an RBC as illustrated in Fig.~\ref{fig:tweezer}.
A stationary state is reached after some time when the tensions in the membrane balance the external forces.
We then obtain the longitudinal and transversal diameters of the stretched RBC.
The physical parameters used for these simulations are $\mods = 5.3\, \upmu \mathrm{N} / \mathrm{m}$ (shear modulus) and $\modb = 2.0 \times 10^{-19}\, \mathrm{N}\, \mathrm{m}$ (bending modulus) which correspond to healthy RBCs.~\cite{suresh_connections_2005, gompper_lipid_2008}
The results as shown in Fig.~\ref{fig:tweezer} reveal excellent agreement between experiment~\cite{suresh_connections_2005} and our simulations.
We conclude that the model can reliably predict RBC deformations up to stretching forces of about $100\, \mathrm{pN}$.

\subsection{Production simulations}

We have performed simulations for a single RBC in DLD devices with various row shifts between $\epsilon = 5/80$ and $22/80$ ($d$ between $2.0$ and $8.8\, \micron$) and capillary numbers between $\Canum = 0.1$ and $1.5$.
For healthy RBCs with $\mods = 5.3\, \upmu \mathrm{N} / \mathrm{m}$ and the chosen DLD device geometry, this corresponds to pressure gradients between $p' = 0.29$ and $4.3\, \mathrm{bar} / \mathrm{m}$ and average flow speeds between $\bar u = 0.033$ and $0.49\, \mathrm{mm} / \mathrm{s}$.

The simulations provide access to cell trajectories, velocities, deformation details and local stresses which are difficult, if not impossible, to obtain in state-of-the-art experiments.
The results are presented and discussed in the following section.

\section{Results and discussion}
\label{sec:results}

We first characterise the cell trajectories (section~\ref{sec:trajectories}) before we analyse the deformation and apparent diameter of the cells (section~\ref{sec:deformation}).

\subsection{Cell trajectories}
\label{sec:trajectories}

\begin{figure}
 \centering
 \subfloat[$\epsilon = 5/80$, $d = 2.0\, \micron$]{\includegraphics[width=\linewidth]{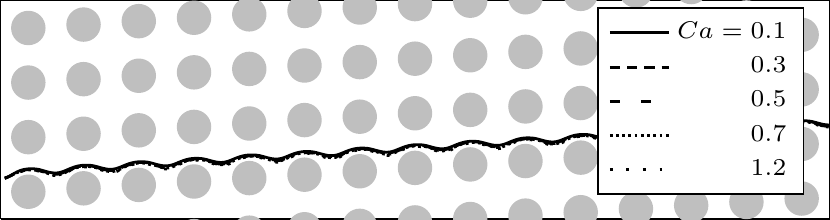}} \\
 \subfloat[$\epsilon = 11/80$, $d = 4.4\, \micron$] {\includegraphics[width=\linewidth]{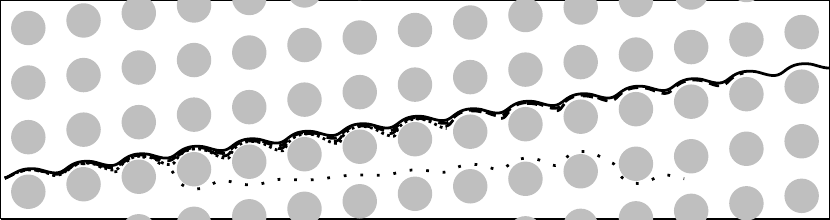}} \\
 \subfloat[\label{fig:trajectories_60} $\epsilon = 15/80$, $d = 6.0\, \micron$] {\includegraphics[width=\linewidth]{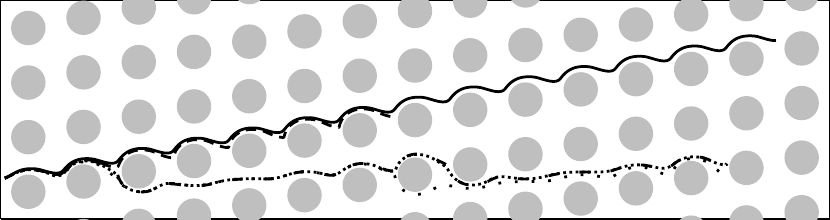}} \\
 \subfloat[$\epsilon = 17/80$, $d = 6.8\, \micron$] {\includegraphics[width=\linewidth]{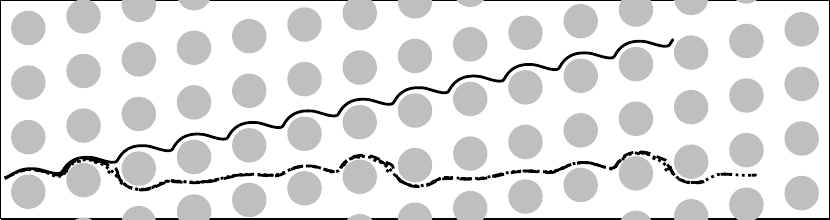}} \\
 \subfloat[\label{fig:trajectories_88} $\epsilon = 22/80$, $d = 8.8\, \micron$] {\includegraphics[width=\linewidth]{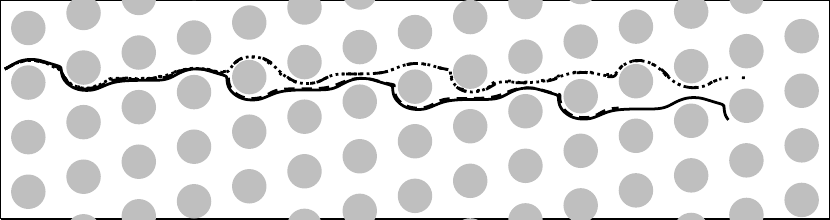}} \\
 \caption{\label{fig:trajectories} RBC trajectories for five different capillary numbers ($\Canum = 0.1$, $0.3$, $0.5$, $0.7$ and $1.2$ as indicated by solid, dashed and dotted lines in the legend of (a)) in five geometries with different row shifts ($\epsilon = 5/80$, $11/80$, $15/80$, $17/80$ and $22/80$ according to $d = 2.0$, $4.4$, $6.0$, $6.8$ and $8.8\, \micron$). For a small row shift ($\epsilon = 5/80$) in (a), all cells are displaced. Increasing the shift (up to $\epsilon = 17/80$) in (b)--(d) initially leads to the zigzag motion of the most supple and later also of more rigid particles. For the largest shift ($\epsilon = 22/80$) in (e), all particles move on zigzag trajectories.}
\end{figure}

In Fig.~\ref{fig:trajectories} we show some trajectories of RBCs for different row shifts $d$ and \Canum-values.
Two extreme cases can be observed.
First, when the row shift is small ($\epsilon = 5/80$), all particles are displaced.
Secondly, for a much larger displacement ($\epsilon = 20/80$), all particles move horizontally on average (with a zigzag motion).
However, one recognises that intermediate row shifts lead to a deformability-dependent displacement.
For example, only the softest RBCs ($Ca > 1$) show zigzag motion for $\epsilon = 11/80$, whereas for $\epsilon = 13/80$ all RBCs with $Ca \geq 0.5$ move on zigzag trajectories.
For $\epsilon = 17/80$ only the most rigid RBC ($Ca = 0.1$) is displaced.

\begin{figure}
 \includegraphics[width=\linewidth]{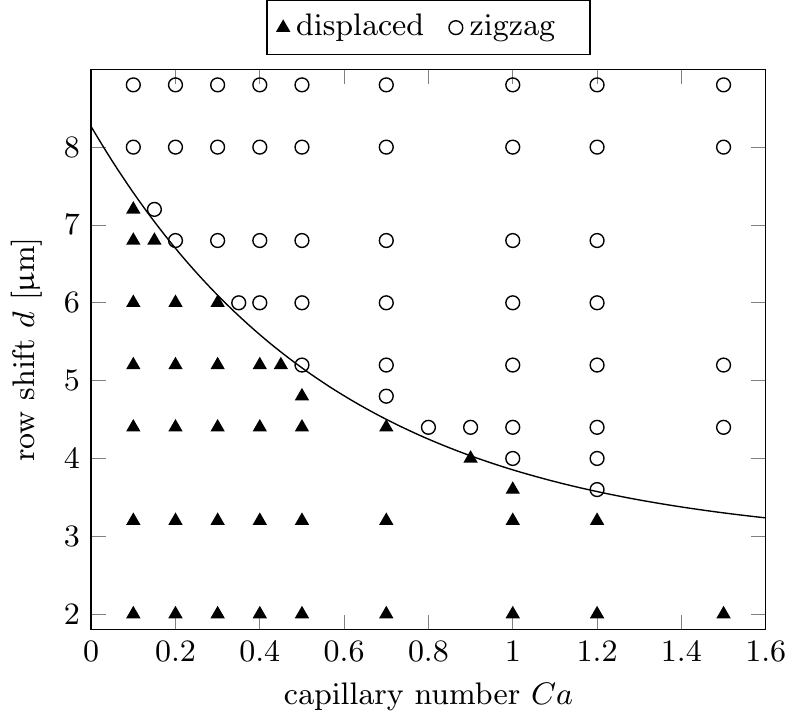}
 \caption{\label{fig:phasespace} Separation characteristics in deformability-row shift space. Solid triangles indicate displaced cells, open circles correspond to cells on zigzag trajectories, \ie cells moving horizontally on average. The solid line corresponds to the function in Eq.~\eqref{eq:phase_boundary}, which is an excellent approximation for the boundary between displaced and zigzag trajectories.}
\end{figure}

The displacement ``phase space'' is shown in Fig.~\ref{fig:phasespace}.
It characterises the shape of the trajectories (displaced or zigzag) as a function of chosen row shift $d$ and capillary number \Canum.
In particular, this figure reveals how $d$ has to be chosen in order to separate particles below and above a specified \Canum threshold.
The boundary between displaced and zigzag trajectories can be very well approximated by the simple exponential function
\begin{equation}
 \label{eq:phase_boundary}
 \frac{d(\Canum)}{\micron} = 2.9 + 5.4\, \mathrm{e}^{- 1.72\, \Canum}
\end{equation}
as shown by the solid line in Fig.~\ref{fig:phasespace}.

For example, assuming that two species of RBCs with a factor of three difference in their shear modulus exist in a given sample (which is a typical value for malaria-infected \emph{versus} healthy cells), the pressure gradient could be selected such that the softer and more rigid RBCs yield $\Canum = 0.4$ and $1.2$, respectively.
A row shift between $4.0$ and $5.5\, \micron$ should therefore be chosen to separate these two RBC species.

We would like to point out that \Canum does not only have an effect on the trajectories.
The average velocity of the RBCs also depends on the capillary number, even if the cells follow similar trajectories.
This can be seen in Fig.~\ref{fig:trajectories_60} where the RBC with $\Canum = 0.3$ follows the same path as the cell with $\Canum = 0.1$, but the velocity of the former is 50\% smaller than that of the latter.
The reason is that, when \Canum is close to its critical value, particles experience head-on collisions with the downstream pillar as they essentially follow a streamline which ends in a stagnation point.
Consequently, it takes a long time for the particle to roll either to the left or the right, which in turn reduces the average flow velocity.
In principle, the RBC may even get stuck for a very long time if \Canum is sufficiently finely adjusted.
We discuss this observation more thoroughly in a different work.~\cite{krueger_deformability_2014}

\subsection{Cell deformation and apparent diameter}
\label{sec:deformation}

\begin{figure}
 \centering
 \subfloat[\label{fig:criticaldiametersketch} Definition of streamline size $\dlat$]{\includegraphics[width=0.5\linewidth]{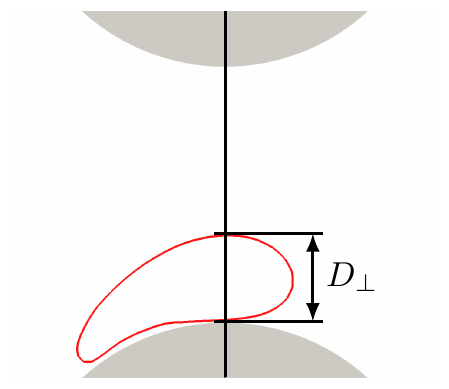}} \\
 \subfloat[\label{fig:critical_diameters015} $\Canum = 0.15$, $\dlat = 6.7\, \micron$]{\includegraphics[width=0.475\linewidth,clip=true,trim=0 0 0 3cm]{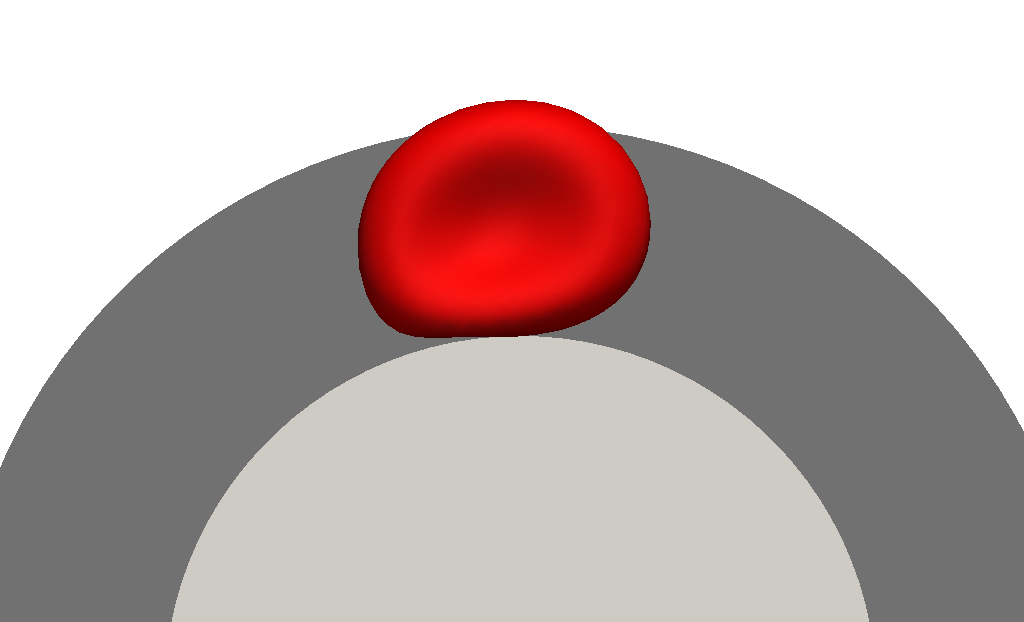}} \hfill
 \subfloat[\label{fig:critical_diameters030} $\Canum = 0.3$, $\dlat = 5.8\, \micron$] {\includegraphics[width=0.475\linewidth,clip=true,trim=0 0 0 3cm]{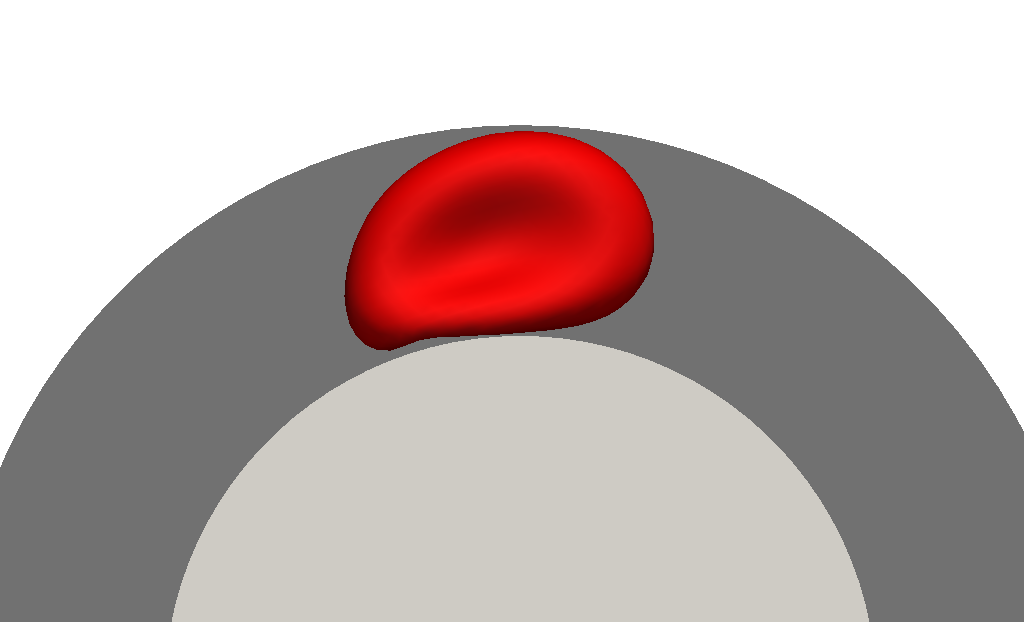}} \\
 \subfloat[\label{fig:critical_diameters070} $\Canum = 0.7$, $\dlat = 4.6\, \micron$] {\includegraphics[width=0.475\linewidth,clip=true,trim=0 0 0 3cm]{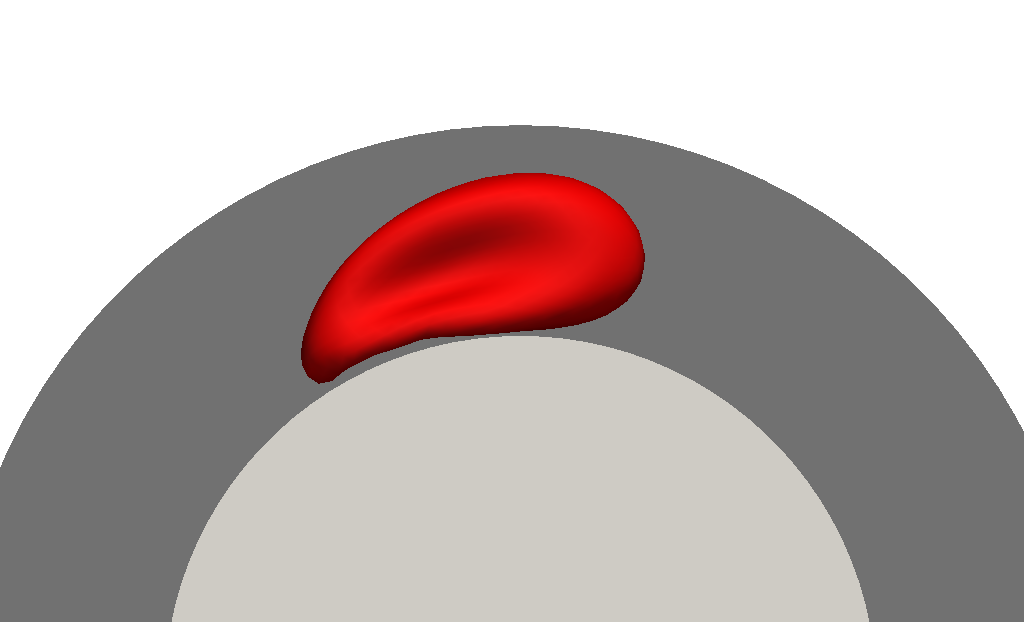}} \hfill
 \subfloat[\label{fig:critical_diameters120} $\Canum = 1.2$, $\dlat = 3.8\, \micron$] {\includegraphics[width=0.475\linewidth,clip=true,trim=0 0 0 3cm]{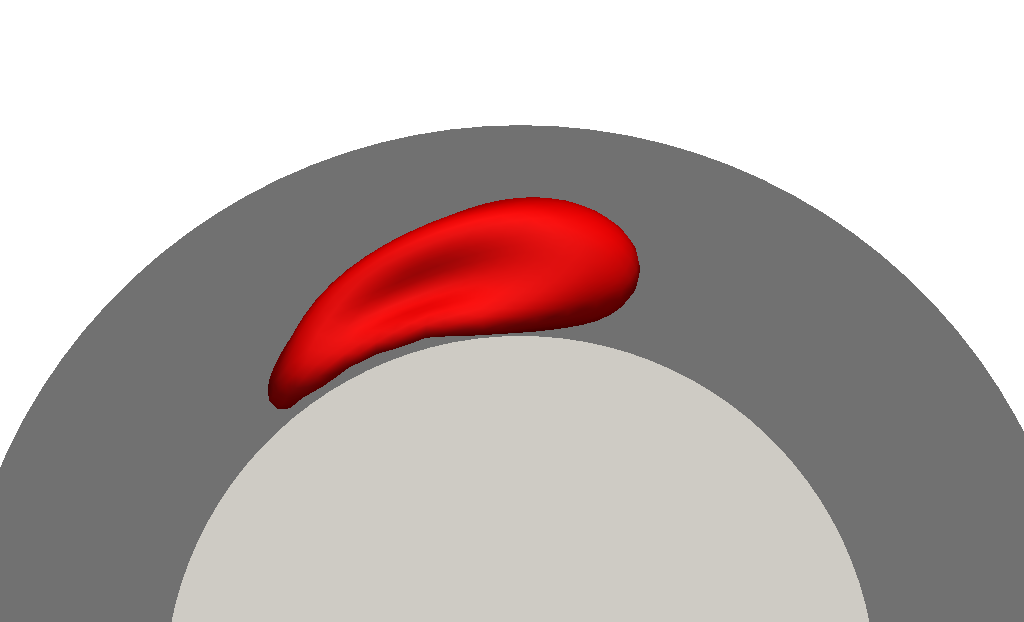}} \\
 \caption{\label{fig:critical_diameters} (Colour online) Effect of deformability on lateral RBC extension. The flow is from left to right. (a) The streamline size $\dlat$ of the RBC (cross-section shown as red line) is defined as its maximum lateral extension during its passage between two pillars (grey circle segments).
 (b)--(e) Particle shapes during passage around the pillar (light grey) for $d = 6.8\, \micron$ ($\epsilon = 0.21$) and different capillary numbers \Canum at the moment when the perpendicular extension $\dlat$ is maximum. For this particular row shift $d$, the critical separation diameter is $2 s = 5.9\, \micron$, as illustrated by the dark grey region. Only the cell in (b) is laterally displaced; the others in (c)--(e) have sufficiently small perpendicular extensions to follow the zigzag trajectories.}
\end{figure}

We obtained Eq.~\eqref{eq:separation_law} based on the idea that particle trajectories are controlled by the shape of the streamlines of the ambient fluid.
We now assume that the separation characteristics of the RBCs are dominated by their instantaneous size during passage between two pillars.
For this reason, we define the \emph{streamline size} \dlat of an RBC as its largest lateral extension along the line connecting the neighbouring pillars the RBC is currently passing, as illustrated in Fig.~\ref{fig:criticaldiametersketch}.

Due to the expected dependence of \dlat on \Canum, RBCs at various values of \Canum should experience different streamlines and may therefore follow different trajectories in the same geometry.
We show the configuration of an RBC during its passage between neighbouring pillars for four different capillary numbers in Fig.~\ref{fig:critical_diameters}(b)--(e).
The row shift is $d = 6.8\, \micron$.
We observe that \dlat decreases upon an increase of $\Canum$.
Initially, for small \Canum in Fig.~\ref{fig:critical_diameters015}, the RBC size is larger than the critical diameter of a sphere in this geometry ($\Dcrit = 5.9\, \micron$, Eq.~\eqref{eq:separation_law}).
In fact, this RBC turns out to be displaced.
For the remaining examples in Fig.~\ref{fig:critical_diameters}(c)--(e), \dlat is smaller than \Dcrit, and those RBCs actually follow the zigzag trajectories.

\begin{figure}
 \includegraphics[width=\linewidth]{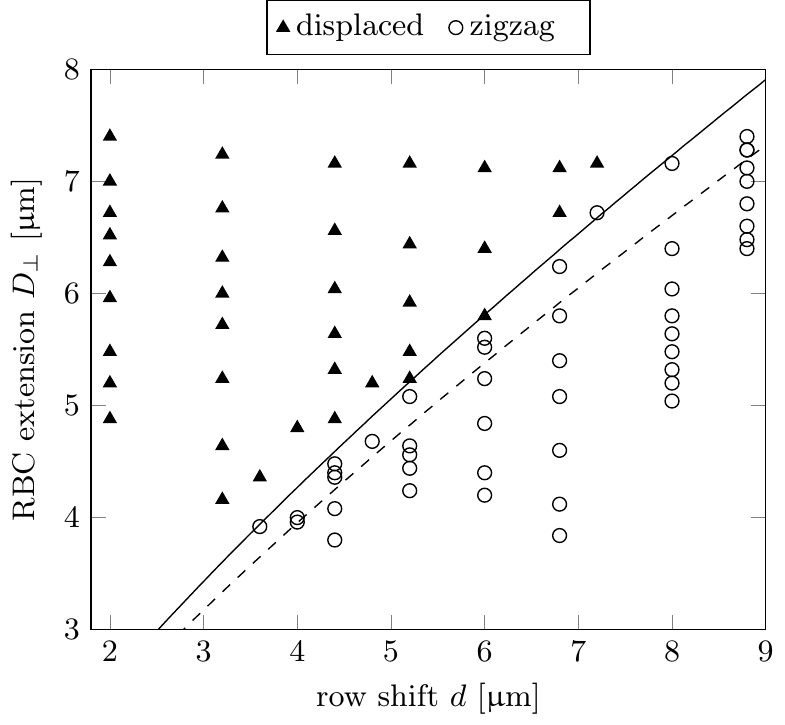}
 \caption{\label{fig:criticaldiameter} Separation characteristics in row shift-RBC extension ($d$-$\dlat$) space. The symbols are the same as in Fig.~\ref{fig:phasespace}. The dashed line corresponds to the function in Eq.~\eqref{eq:separation_law} while the solid line is the same function multiplied by a correction factor of $1.08$.}
\end{figure}

We have analysed the shapes of all RBCs during their passage between the pillars.
Fig.~\ref{fig:criticaldiameter} shows $\dlat$ for each RBC as a function of row shift $d$.
Depending on the capillary number, there are multiple values for each row shift where \dlat generally decreases for increasing \Canum.
One can clearly see a sharp separation between the displaced and the zigzag trajectories, which can be described relatively well by the empirical law in Eq.~\eqref{eq:separation_law} (dashed line).
This supports the above idea that the extension of a deformed RBC at the moment it passes the region of highest streamline density determines which trajectory to follow afterwards.

The prediction can be significantly improved by multiplying Eq.~\eqref{eq:separation_law} by a correction factor of $1.08$ (solid line in Fig.~\ref{fig:criticaldiameter}).
It is not surprising that the actual separation characteristics are not perfectly captured by Eq.~\eqref{eq:separation_law}.
First, the particles are extended in the third dimension as well.
Secondly, the presence of the RBC changes the shape of the streamlines compared to the situation without particles, which in turn has an effect on the motion of the RBC.
This effect is expected to be significantly more important for denser RBC suspensions.

\begin{figure}
 \includegraphics[width=\linewidth]{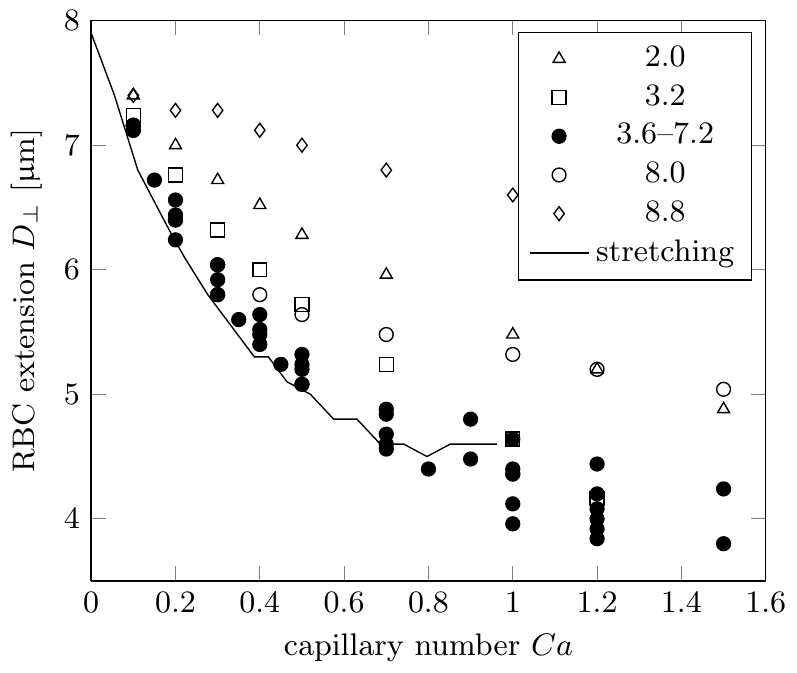}
 \caption{\label{fig:deformation} Cell extension \dlat during passage between two pillars as a function of capillary number \Canum. Data for relevant row shifts ($d \in [3.6, 7.2]$) are shown as filled symbols, the remaining data are shown as open symbols. The simulated small cell diameters from Fig.~\ref{fig:tweezer} are plotted as a solid line where the stretching force $F$ has been converted to a capillary number according to Eq.~(\ref{eq:capillary_conversion}).}
\end{figure}

We now investigate the deformation behaviour of the RBCs as a function of the capillary number.
Fig.~\ref{fig:deformation} shows the same data as Fig.~\ref{fig:criticaldiameter}, but plotted \emph{versus} \Canum rather than row shift $d$.
The first observation is that all data for $d \in [3.6, 7.2] \micron$ basically collapse on a single line $\dlat(\Canum)$ (filled symbols).
RBC sizes for small ($d \leq 3.2\, \micron$) or large ($d \geq 8.0\, \micron$) row shifts do not follow this trend (open symbols).
Interestingly, these extreme cases are those for which no \Canum-dependent trajectories have been observed (Fig.~\ref{fig:phasespace}).
The reason is that for $d \leq 3.2\, \micron$, the critical diameter is so small that even strongly deformed RBC are large enough to be displaced.
Similarly, for $d \geq 8.0\, \micron$, even nearly rigid RBCs are too small to be displaced; they all follow zigzag trajectories.
Therefore, only intermediate row shifts are suitable for a \Canum-dependent separation of RBCs anyway.
We will therefore exclude the extreme row shifts (all open symbols in Fig.~\ref{fig:deformation}) from the following discussion.
We note that the row shift range for which separation is possible will generally depend on other device parameters such as pillar shape and size.
We also find that those RBCs which are shown with open symbols in Fig.~\ref{fig:deformation} do not come in close contact with pillars during their passage, \ie there is always an appreciable fluid layer between the RBC and the pillar.
Due to the absence of direct collisions with the pillar, the deformation characteristics are different so that the $\dlat(\Canum)$ curves do not collapse on the master curve.
It is therefore possible that direct collisions of RBCs with obstacles actually increase the separation sensitivity; this should be taken into account in future DLD designs.

RBCs in geometries with $d \in [3.6, 7.2] \micron$ are in close contact with the pillar during their passage.
The cells therefore move through a region of large viscous stresses.
A simple shear flow can be represented as a superposition of a rotational and an extensional component.
We now hypothesise that the extensional stress during the passage between the pillars has qualitatively the same effect as the stretching force in an optical tweezer (Fig.~\ref{fig:tweezer}).
The rotational component is suppressed by the presence of the pillar as a rigid obstacle.

We define a \emph{stretching capillary number} for the optical tweezer setup:
\begin{equation}
 \label{eq:capillary_conversion}
 \Canum_\mathrm{tw} = \frac{\sigma_\mathrm{tw} r}{\mods}, \quad \sigma_\mathrm{tw} = \frac{2 F}{A},
\end{equation}
where $2 F$ is the total stretching force acting on the RBC surface area $A$ (which is $8.72 r^2$ for an RBC).
Associated with this is the average surface stretching stress scale $\sigma_\mathrm{tw} = 2 F / A$.
The simulation data of the small RBC diameter from Fig.~\ref{fig:tweezer_data} can now be plotted as a function of $\Canum_\mathrm{tw}$ rather than $F$.
It is shown as a solid line in Fig.~\ref{fig:deformation}.
Note that there are no free parameters involved.
The agreement of the stretching data and the simulated RBC extensions in the DLD device is remarkable.

We can therefore conclude that the stretching component of the viscous stress in the DLD geometry has qualitatively and quantitatively a similar effect on RBC deformation as the stretching force in an optical tweezer.
In particular the extension of the RBC along its shortest diameter behaves nearly identically, at least for those RBCs which come in close contact during their passage between pillars ($d \in [3.6, 7.2] \micron$).
This is interesting because the actual RBC shape is not exactly the same in both situations.

Based on these findings, it is possible to predict for which row shift $d$ and capillary number \Canum an RBC will be displaced or follow the zigzag trajectories. For this, we need
\begin{enumerate}
 \item the known stretching curve in Fig.~\ref{fig:tweezer_data} to relate the small RBC diameter to the stretching force and therefore to the capillary number,
 \item the critical diameter \Dcrit for a given DLD device which can be simply obtained by investigating the shape of the streamlines or performing experiments with rigid beads.
\end{enumerate}

Our observations and results should help in the design of cell separation devices beyond simple trial and error.
However, we emphasise again that the results reported here are only valid for a specific DLD device, in particular having a height of $4.8\, \micron$.
The ability to better understand and accurately design DLD structures should lead to an expansion in their functionality and wider use for separation of biological mixtures of ever increasing complexity.

\section{Conclusions}

We have performed three-dimensional high resolution immersed-boundary-lattice-Boltzmann-finite-element simulations of single deformable red blood cells (RBCs) in deterministic lateral displacement (DLD) devices.
While keeping other geometrical parameters fixed, we have varied the row shift $d$ of the DLD setup.
Additionally we have varied the RBC capillary number \Canum, the ratio between deforming viscous stresses and restoring elastic stresses of the RBC membrane.
For example, more rigid RBCs have a smaller \Canum value than suppler cells in the same flow environment.
Inertial effects are negligible in the current situation.

Investigating the cell trajectories and displacement characteristics in the $\Canum$-$d$ diagram, we have shown that a deformability-based separation of RBCs in DLD devices is possible.
We observe displaced trajectories, where the cells move diagonally on average, and zigzag trajectories, where the RBCs basically follow horizontal streamlines.
In the past, RBC deformation was often considered an undesired effect in DLD geometries.
On the contrary, our work shows how to take advantage of RBC deformation.
For example, it could be used to separate early stage malaria-infected RBCs from healthy ones since the former are usually more rigid than the latter, although both have nearly identical undeformed shapes.

We have further identified the instantaneous lateral RBC extension as a key parameter for cell separation.
Since the RBC size depends on \Canum, for each row shift $d$ there is a critical capillary number above which the RBC appears sufficiently small to follow the zigzag trajectories and below which it appears so large that it is laterally displaced.
For the particular DLD device modelled here, we identified the relevant range of row shifts for RBC separation to be between $3.2$ and $8.0\, \micron$.
Below $d = 3.2\, \micron$, all investigated RBCs are displaced and above $d = 8.0\, \micron$, all move on zigzag trajectories.

Additionally we have rationalised the lateral RBC extension, which eventually determines the fate of the cell trajectories, in terms of the extensional component of the shear flow in the region between the pillars.
We showed that the lateral RBC extension behaves similarly for the same level of extensional stress, independent of whether this stress is caused by an optical tweezer or the viscous flow in the DLD device.
This may help to predict the trajectories of RBCs in hitherto untested DLD geometries and therefore reduce the high level of trial and error involved in their design and calculation.

Our work should therefore help in the design of new DLD setups for deformability-based RBC separation and to understand the trajectories of RBCs in such devices.
This may eventually facilitate the design of cheaper, faster and more robust diagnostic devices for the detection of malaria and other diseases.

\section*{Acknowledgements}

This work was supported by the EPSRC grant ``Large Scale Lattice Boltzmann for Biocolloidal Systems'' (Grant No. EP/I034602/1) and the EC-FP7 project ``CRESTA'' (\url{http://www.cresta-project.eu/}; Grant No. 287703).
TK thanks the University of Edinburgh for the award of a Chancellor's Fellowship and Dr Prashant Valluri for stimulating discussions.
DH acknowledges the award of a UCL EPSRC Strategic Enterprise Award (GR/T11364/01) and thanks Prof Gabriel Aeppli and the late Prof Tom Duke for their support while at the London Centre for Nanotechnology.

\balance

\footnotesize{
\providecommand*{\mcitethebibliography}{\thebibliography}
\csname @ifundefined\endcsname{endmcitethebibliography}
{\let\endmcitethebibliography\endthebibliography}{}

}

\end{document}